\documentclass[12pt,preprint]{aastex}
\newcommand{\eso}{ESO\,338$-$IG04B}
\newcommand{\etal}{et\,al.}
\newcommand{\gsim}{\raise0.3ex\hbox{$>$}\kern-0.75em{\lower0.65ex\hbox{$\sim$}}}
\newcommand{\halpha}{H$\alpha$}
\newcommand{\iras}{IRAS\,08339+6517}
\newcommand{\kms}{km\,s$^{-1}$}
\newcommand{\jkms}{\,\,Jy\,km\,s$^{-1}$}
\newcommand{\lya}{Ly$\alpha$}
\newcommand{\mjpbeam}{\,\,mJy\,Beam$^{-1}$}
\newcommand{\msun}{M$_{\odot}$}
\newcommand{\HI}{H~{\sc i}}

\newcommand{\tol}{Tol\,1924$-$416}
\newcommand{\twomass}{2MASX\,J08380769+6508579}
\begin{document}     
\title{Extended Tidal Structure In Two \lya-Emitting Starburst Galaxies}
\author{John M. Cannon and Evan D. Skillman}
\affil{Department of Astronomy, University of Minnesota,\\ 116 Church St. 
S.E., Minneapolis, MN 55455}
\email{cannon@astro.umn.edu, skillman@astro.umn.edu}
\author{Daniel Kunth}
\affil{Institut d'Astrophysique,\\ Paris, 98bis Bld Arago, F-75014, 
Paris, France}
\email{kunth@iap.fr} 
\author{Claus Leitherer}
\affil{Space Telescope Science Institute,\\ 3700 San Martin Drive, 
Baltimore, MD 21218}
\email{leitherer@stsci.edu} 
\author{Miguel Mas-Hesse}
\affil{Centro de Astrobiologia (CSIC-INTA),\\ E-28850 Torrejon de Ardoz, 
Madrid, Spain}
\email{mm@laeff.esa.es} 
\author{G{\" o}ran {\" O}stlin}
\affil{Stockholm Observatory,\\ SE-106 91 Stockholm, Sweden}
\email{ostlin@astro.su.se} 
\author{Artashes Petrosian}
\affil{Byurakan Astrophysical Observatory and Isaac Newton Institute of 
Chile,\\ Armenian Branch, Byurakan 378433, Armenia}
\email{artptrs@yahoo.com}
\begin{abstract}

We present new VLA C-configuration \HI\ imaging of the \lya-emitting starburst 
galaxies \tol\ and \iras.  The effective resolution probes neutral gas 
structures larger than 4.7 kpc in \tol, and larger than 8.1 kpc in \iras.  
Both systems are revealed to be tidally interacting:  \tol\ with \eso\ 
(6.6\arcmin\ = 72 kpc minimum separation), and \iras\ with \twomass\ 
(2.4\arcmin\ = 56 kpc minimum separation).  The \HI\ emission is extended in 
these systems, with tidal tails and debris between the target galaxies and 
their companions.  Since \lya\ emission has been detected from both of these 
primary systems, these observations suggest that the geometry of the ISM is 
one of the factors affecting the escape fraction of \lya\ emission from 
starburst environments.  Furthermore, these observations argue for the 
importance of interactions in triggering massive star formation events.  

\end{abstract}						


\section{Introduction}
\label{S1}

Here, we present new VLA \HI\ imaging of the starburst galaxies \tol\
[$\alpha$,$\delta$ (J2000) $=$ 19:27:58.2, $-$41:34:32; D = 37.5 Mpc {({\" 
O}stlin, Bergvall, \&\ Roennback 1998}\nocite{ostlin98})] and \iras\ 
[$\alpha$,$\delta$ (J2000) $=$ 08:38:23.2, 65:07:15; D $\simeq$ 80 Mpc 
(Assuming H$_0$ = 72 km\,sec$^{-1}$\,Mpc$^{-1}$; {Freedman \etal\ 
2001}\nocite{freedman01})].  Both galaxies reveal extended tidal structure 
in neutral hydrogen and are clearly interacting with nearby companions.  
Each appears to provide evidence for tidally-induced starburst episodes.  

These two systems were chosen for imaging because they exhibit prominent 
\lya\ emission \citep{leitherer95,giavalisco96,kunth98,leitherer02}. Hydrogen
\lya\ is one of the most important diagnostic emission lines in astrophysics.  
It is predicted to be luminous in star-forming galaxies \citep{charlot93}, 
and could potentially be used as an indicator of star formation activity in 
high-redshift systems.  However, this application is not straightforward, 
since the propagation of \lya\ photons in starburst galaxies appears to be a 
complex process that depends on many factors intrinsic to both the starburst 
itself (e.g., star formation rate, metallicity) and to the surrounding galaxy 
(e.g., geometry, kinematics, and dust content).

In a large spectroscopic sample of comparatively high-redshift Lyman break 
galaxies (LBGs; these systems have typical star formation rates $\gsim$ 10
\msun\,yr$^{-1}$ and are thus directly comparable to more local ``starburst''
galaxies), \citet{shapley03} find various correlations between the emergent 
\lya\ profile (i.e., strengths of absorption or emission) and characteristics 
of the ISM.  In particular, dustier galaxies have less prominent \lya\ 
emission, or equivalently, redder UV continuum slopes.  \lya\ emission 
strength is shown to anti-correlate with the kinematic offset implied by the 
redshift difference between \lya\ emission and low-ionization interstellar 
absorption lines; that is, systems with larger outflow speeds usually show 
less prominent \lya\ emission.  Finally, systems with stronger \lya\ emission 
show lower star formation rates, on average, than systems with weaker \lya\
emission.  These correlations suggest that many factors can contribute to the 
escape fraction of \lya\ from starburst regions.  

One of the most important factors that influences \lya\ propagation is resonant
scattering.  It is well known that due to high resonant scattering by neutral
hydrogen in the ISM, \lya\ photons can be attenuated by even small amounts of 
dust.  It is then expected that only young, relatively dust-free galaxies 
should be prodigious sites of \lya\ production.  Low-metallicity starburst 
galaxies in the local universe may be considered nearby analogs to such 
objects which are expected in greater numbers at higher redshifts.  Thus it 
was surprising that HST observations of the most metal-poor galaxy known, 
I\,Zw\,18, showed only damped \lya\ absorption and no emission 
\citep{kunth94}.  In stark contrast, the more metal-rich starburst galaxy 
Haro\,2 showed prominent \lya\ emission \citep{lequeux95}.  

The effects of resonant scattering can be lessened if there is a velocity
difference between the site of \lya\ emission and the surrounding material 
from which resonant scattering can occur.  This will Doppler shift the \lya\ 
photon out of resonance and increase its chance of escaping the surrounding 
ISM.  While this process seems to be complicated (note, for example, that 
higher outflow speeds suggest less prominent \lya\ emission profiles in the 
sample of {Shapley \etal\ 2003}\nocite{shapley03}), it appears that an 
important factor is the presence or absence of an outflow (see below). Indeed,
\citet{shapley03} measure an average \lya\ offset $>$ 300\,\kms\ for the 
sample of \lya-emitting LBGs.  These observations and the studies of 
{Giavalisco \etal\ (1996)}\nocite{giavalisco96} and \citet{kunth98} argue for 
the importance of ISM geometry and kinematics in affecting the \lya\ escape 
fraction.  

The studies of \citet{kunth98}, \citet{tenoriotagle99}, and \citet{mashesse03}
have examined the characteristics of the ISM kinematics which appear to 
influence the propagation of \lya\ photons in star-forming galaxies.  If 
static, homogeneous neutral gas with column densities $\gsim$ 
10$^{18}$\,cm$^{-2}$ shields the ionized gas, no emission will be detected.  
The resonant scattering of the \lya\ photons will lead to increased 
probability of destruction by any dust which is present.   On the other hand, 
there may be diffuse \lya\ emission which is detectable on sightlines not 
coincident with the sources of UV photons.  Similarly, if the areal coverage 
of the neutral gas is not uniform but clumpy, some \lya\ emission may be 
detectable on favorable sightlines.  Finally, if the velocity structure of the
neutral gas is not static but rather outflowing from the ionizing regions 
(outflow velocities \gsim\,200\,\kms),  \lya\ photons to the red of 1216\,\AA\
can escape and \lya\ emission may be significant.  This explains the strong 
\lya\ emission detected in some starburst galaxies with complete spatial 
coverage by neutral gas which is also comparatively rich in both metals and 
dust.

These observations were undertaken with these arguments in mind.  Our group has
obtained new HST/ACS imaging of the \lya\ line in a small sample of galaxies
(see {Kunth \etal\ 2003}\nocite{kunth03}), two of which are \tol\ and \iras.  
While these data are only a small segment of the sample that was imaged in 
\lya, they will aide in the interpretation of the \lya\ data.  We note that 
our \HI\ imaging sample does not include systems showing only \lya\ 
absorption, so we must await a larger sample to draw definitive conclusions on
the correlation between \HI\ kinematics and the appearance of \lya\ emission.  
However, the results presented here support the scenario that the geometry and
kinematics of the ISM are important (although not the only) factors that 
affect the escape probability of \lya\ photons from massive star formation 
regions.

\section{Observations and Data Reduction}
\label{S2}

The NRAO Very Large Array\footnote{The National Radio Astronomy Observatory is 
a facility of the National Science Foundation operated under cooperative 
agreement by Associated Universities, Inc.} (VLA) was used to obtain \HI\ 
spectral line data for these two systems, as part of observing program 
AC\,654.  \tol\ was observed in the BnC configuration for 180 minutes on 2002 
September 30, while \iras\ was observed in the C configuration for 370 minutes
on 2002 December 8 \&\ 9.  The correlator configuration yields a bandwidth of 
3.125 MHz, with 64 channels separated by 48.8 kHz (10.5 \kms).  The data were 
reduced and calibrated using standard methods in the AIPS environment.  The 
data were imaged using uniform weighting; the single-plane rms noise is 1.3 
\mjpbeam\ for \tol\ and 0.40 \mjpbeam\ for \iras.  With the final beam sizes 
(26\arcsec$\times$16\arcsec\ for \tol, and 21\arcsec$\times$16\arcsec\ for 
\iras), this noise level implies single-plane 3$\sigma$ detection thresholds 
of 2.3$\times$10$^{20}$ cm$^{-2}$ and 6.9$\times$10$^{19}$ cm$^{-2}$ for \tol\
and \iras, respectively. 

\section{\HI\ Data: Extended Tidal Structure}
\label{S3}

With these data we detect extended tidal structure in each of these systems.
In both cases \HI\ gas is detected in the target system, in a neighboring 
galaxy, and in tidal material.

\subsection{\tol}
\label{S3.1}

\begin{figure}
\plotone{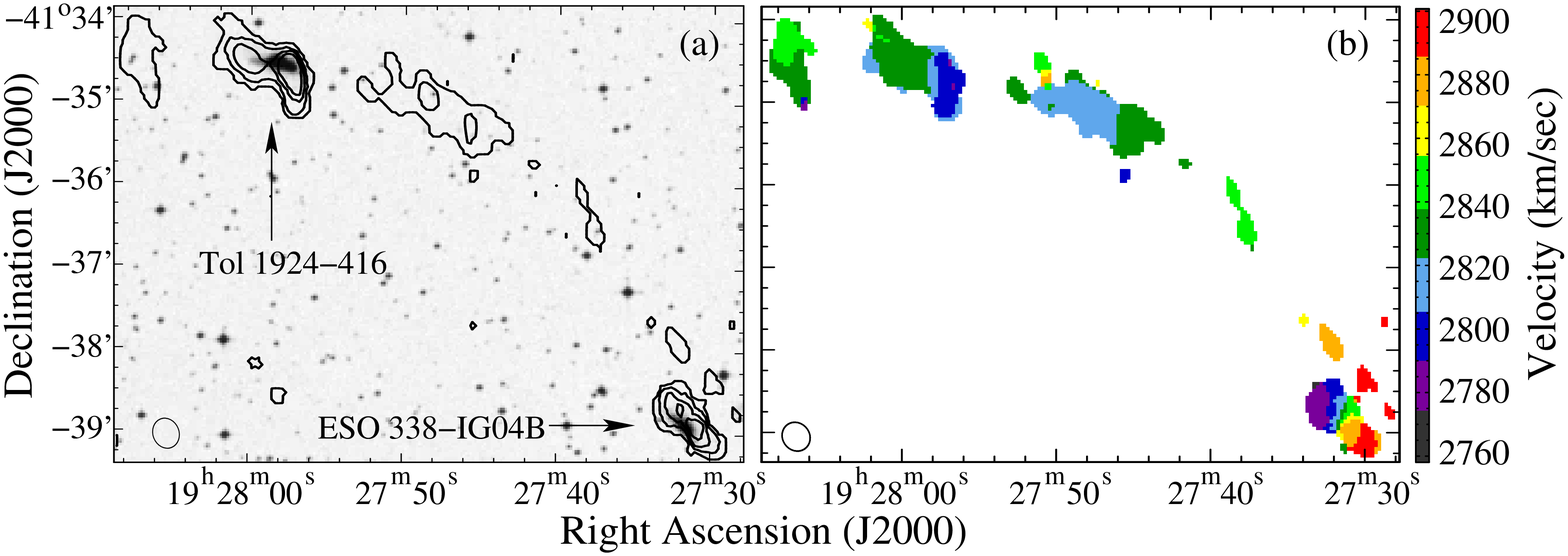}
\caption{(a) DSS image of \tol, overlaid with contours of the \HI\ 
zeroth-moment image.  Contours correspond to column densities of (7.3, 29, 51, 
73)$\times$10$^{20}$ cm$^{-2}$.  Each galaxy is labeled; beam size is shown at 
bottom left. (b) Intensity-weighted velocity field of \tol.  From this figure 
it is apparent that \HI\ is being removed from one or both systems.  Also, 
there remains a component of solid-body rotation within the optical extent of 
both galaxies.  Beam size is labeled at lower left.}
\label{figcap1}
\end{figure}

In Figure~\ref{figcap1} we present contours of the zeroth-moment (total column 
density) image overlaid on a DSS optical image of \tol, as well as an 
intensity-weighted velocity field.  We detect a total integrated flux of 
12.6$\pm$1.58\jkms\ from the \tol\ system.  This compares well with the 
single-dish value of 11.4\jkms\ derived from Parkes observations in the HIPASS 
Survey\footnote{The Parkes telescope is part of the Australia Telescope which 
is funded by the Commonwealth of Australia for operation as a National 
Facility managed by CSIRO; see http://www.atnf.csiro.au/research/multibeam/} 
\citep{barnes01}, suggesting that we have missed little or no flux due to the 
lack of zero-spacing interferometric elements. At the adopted distance of 
37.5 Mpc, this corresponds to a total \HI\ mass of 
(4.2$\pm$0.5)$\times$10$^9$ \msun.  

Inspecting the \HI\ data cube allows us to measure the masses of neutral gas 
clearly associated with both \tol\ and with the companion \eso.  In the former
we detect a total mass of (1.4$\pm$0.2)$\times$10$^9$ \msun, and in the latter
we find (9.3$\pm$1.2)$\times$10$^8$ \msun.  The remainder of the system mass,
(1.9$\pm$0.2)$\times$10$^9$ \msun\ or $\sim$ 40\%, is tidal material (see 
Figure~\ref{figcap1}). 

The velocity structures of \tol\ and \eso\ have been studied in the \halpha\
line by \citet{ostlin99} and \citet{ostlin01b}.  The agreement between the 
optical and radio velocity fields is very good.  The radio data confirms the 
weak large-scale ordered rotation in \tol, and the very ordered rotation in 
the companion system.  Note that the \HI\ associated with the \lya-emitting 
system (\tol) is spread out over $\sim$ 80 \kms\ in radial velocity, 
suggesting that there is not a static screen of neutral gas along the line of 
sight to the starburst region. 

\tol\ shows a variety of interesting characteristics in optical imaging.  
\citet{ostlin03} used multicolor HST images to derive the cluster formation 
history.  The elevated star formation rate ($\sim$ few \msun\,yr$^{-1}$) has 
been present for $\sim$ 40 Myr, and shows evidence for propagation throughout 
the disk.  There appears to be little to no reddening in this system, in 
agreement with expectations based on the sample of LBGs in \citet{shapley03}.
While each of these parameters likely influence the propagation of \lya\ 
photons through the ISM, our observations suggest that the ISM kinematics are 
also an important factor in this process.

The angular distance between the two galaxies is $\sim$ 6.6\arcmin, 
corresponding to a minimum deprojected distance of $\sim$ 72 kpc.  The total 
radial velocity extent is $\sim$ 150 \kms\ (V$_{sys}$ $=$ 2830 \kms), but 
components exist in both systems at the same velocities.  Hence it seems 
likely that these two systems may now be gravitationally bound.  Regardless, 
the extended nature of the \HI\ suggests that the present starburst in \tol\ 
likely was triggered by a strong gravitational interaction with \eso, and that 
the different star formation rates in these two otherwise similar systems 
[comparable \HI\ masses (see above) and stellar masses (see {{\" O}stlin 
\etal\ 2001}\nocite{ostlin01b})] are attributable to the recent tidal 
interaction between them.  

\subsection{\iras}
\label{S3.2}

\begin{figure}
\plotone{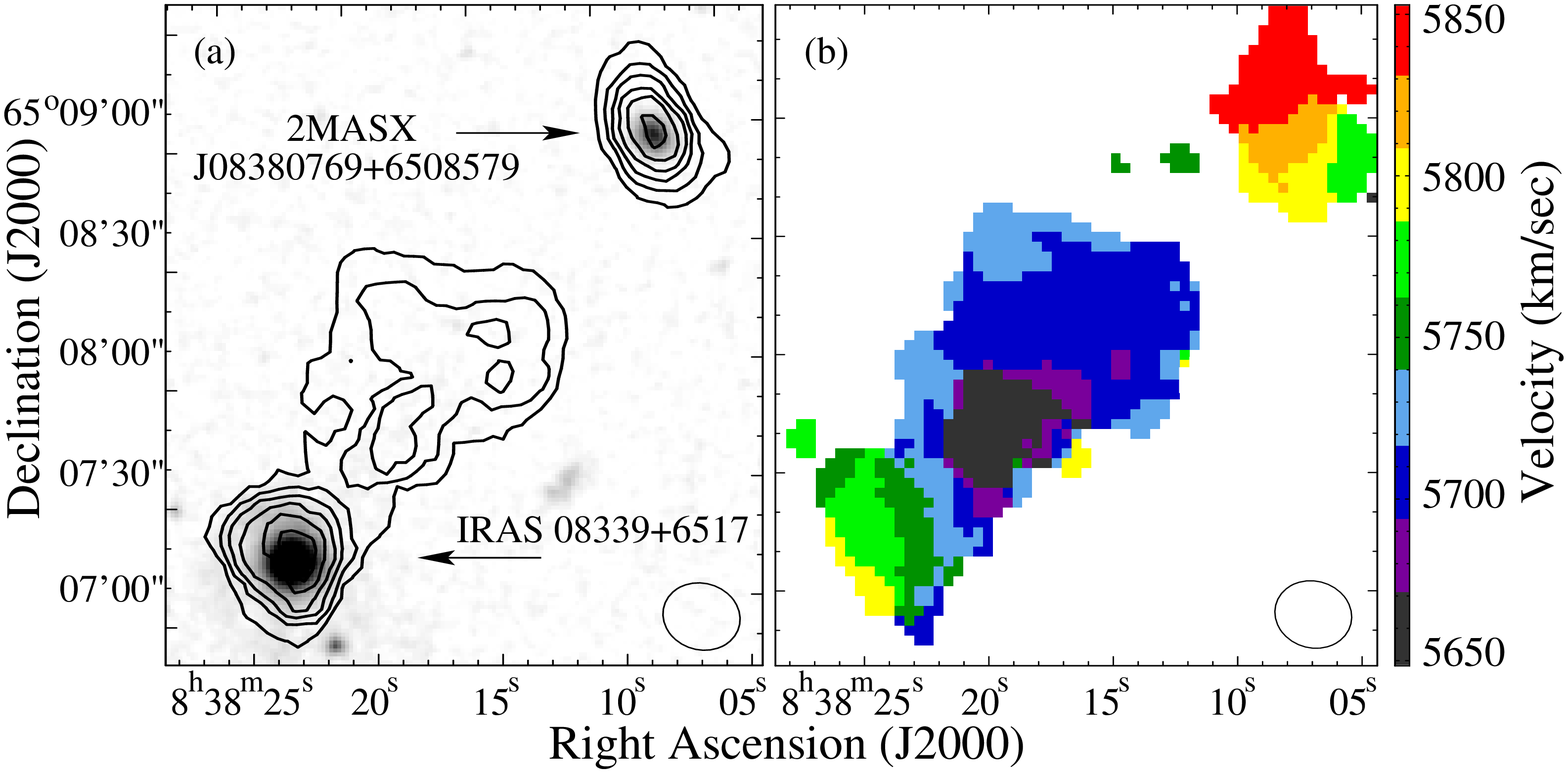}
\caption{(a) DSS image of \iras, overlaid with contours of the \HI\ 
zeroth-moment image.  Contours correspond to column densities of (5.5, 15, 24,
33, 42, 51)$\times$10$^{20}$ cm$^{-2}$.  Each galaxy is labeled; beam size is 
shown at bottom right.  (b) Intensity-weighted velocity field of \iras.  From 
this figure it is apparent that \HI\ is being removed from one or both 
systems.  The companion galaxy appears to retain a component of solid-body 
rotation in neutral gas; clear signs of rotation are less prominent in \iras, 
however, suggesting that this interaction has completely disrupted the neutral
gas in this system. Beam size is labeled at lower right.}
\label{figcap2}
\end{figure}

In Figure~\ref{figcap2} we present contours of the zeroth-moment (total column 
density) image overlaid on a DSS optical image of \iras, as well as an 
intensity-weighted velocity field.  We detect a total integrated flux of 
3.68$\pm$0.46\jkms\ from the \iras\ system; this compares well with the 
single-dish value derived from Nan{\c c}ay observations, 3.81\jkms\ 
\citep{martin91}, again suggesting negligible flux loss due to lack of 
zero-spacing baselines.  At the adopted distance of 80 Mpc, this corresponds 
to a total \HI\ mass of (5.6$\pm$0.7)$\times$10$^9$ \msun.

We find (1.1$\pm$0.2)$\times$10$^9$ \msun\ of \HI\ to be associated with 
\iras.  Its companion system, \twomass, is found to have a total \HI\ content 
of (7.0$\pm$0.9)$\times$10$^8$ \msun, leaving the remaining 
(3.8$\pm$0.5)$\times$10$^9$ \msun\ to reside in tidal material between the 
two systems.  This implies that the bulk of the neutral gas ($\sim$ 70\%) in 
this interacting system has been removed from one or both galaxies.  
Interestingly, the peak single-plane column density is found in tidal material
and is not associated with either galaxy.

\citet{mashesse03} have shown from the analysis of HST/STIS high-resolution UV
spectroscopy that in front of the central $\sim$ 5 kpc (13\arcsec) of this 
galaxy a column density $\sim$ 10$^{19.9}$ cm$^{-2}$ of neutral gas is 
outflowing with a velocity of around $-$300 \kms. This flow of material 
leaving the central area of \iras\ could be feeding the region between both 
galaxies. It is interesting to note that the neutral gas in this intermediate 
region is approaching us with an average velocity of around $-$100 \kms.  
However, no neutral gas is detected in the STIS observations at the systemic 
velocity of the galaxy (V${sys}$ $=$ 5750 \kms) derived from these radio 
observations.  This implies the existence of finer details in the velocity 
structure than can be inferred from these \HI\ observations (e.g., an 
expanding shell powered by the starburst; see {Mas-Hesse \etal\ 
2003}\nocite{mashesse03}).

The distance of \iras\ (80 Mpc) precludes detailed optical studies of the 
stellar or cluster populations.  Inferences about other intrinsic properties of
the galaxy must therefore rely on spectral data.  \citet{gonzalezdelgado98}
measure an \halpha\ flux corresponding to a star formation rate of $\sim$ 8
\msun\,yr$^{-1}$ \citep[applying the prescription of][]{kennicutt94}.  The 
measured reddening reaches values as large as E(B$-$V) $=$ 0.5 mag., suggesting
substantial amounts of dust in this system.  Again we note that these factors 
likely contribute to the propagation of \lya\ photons.  However, since this 
starburst region appears to suffer from substantial reddening, the importance 
of ISM kinematics in Doppler shifting the \lya\ photons out of resonance 
appears to be pronounced.

For a measured angular separation of 2.4\arcmin, the implied minimum 
separation is 56 kpc.  The \HI\ radial velocity extent is $\sim$ 300 \kms;  
this suggests that the neutral gas surrounding \iras\ is highly turbulent, 
having been powered by the release of mechanical energy in the central 
starburst.  At a smaller spatial scale the turbulence is still higher,
as derived from UV spectroscopy \citep[see][]{mashesse03}.

While we can not definitively state that these galaxies now form a bound 
system, the effects of the gravitational interaction are pronounced in \iras.  
Interestingly, this starburst system shows two peaks of \lya\ emission in our 
ACS imaging, and may be the remnant of an earlier merger in its own right. 
On the other hand, the comparatively under-luminous nature of \twomass\ (i.e., 
no obvious massive starburst episode in the current epoch) suggests that the 
effects of the interaction have not been as severe for the secondary system.  
Again, the less-massive system appears to retain a strong component of 
solid-body rotation.  \iras\ also appears to be undergoing organized rotation, 
but the neutral gas is spread out along the line of sight over $\sim$ 80 \kms.

\section{Conclusions} 
\label{S4}

VLA \HI\ imaging of the starburst galaxies \tol\ and \iras\ has been presented.
These two systems are remarkably similar in \HI\ content, mass, and current 
evolutionary state.  In each, we find extended neutral gas between the 
target and nearby neighbors, suggesting that interactions have 
played an important role in triggering the massive starbursts in the primary
galaxies.  The close proximity of the companions suggests that the 
interactions were recent, and the similar velocities of both primary and 
secondary galaxies argues that these systems may end up gravitationally 
bound.

Since both primary systems are intense \lya\ emitters, these data support the 
interpretation that the ISM kinematics are an important mechanism that 
affects the escape of \lya\ photons from starburst regions.  Combined with the
results on LBGs \citep{shapley03}, these data have immediate implications for 
the use of the strength of \lya\ emission in determining star formation rates,
since the results will be dependent on the geometry of the ISM and other 
factors (e.g., dust, star formation rate) and not on properties inherent to 
the starburst being considered.   Further \HI\ observations of \lya-emitting 
galaxies (and, conversely, of starburst systems with no apparent \lya\ 
emission) are certainly warranted to further explore the role of the ISM in 
regulating the escape of \lya\ photons from starburst environments.  

\acknowledgements

Support for this work was provided by NASA through grant number GO-9470 from 
the Space Telescope Science Institute, which is operated by AURA, Inc., under 
NASA contract NAS5-26555.  J.\,M.\,C. is supported by NASA Graduate Student 
Researchers Program (GSRP) Fellowship NGT 5-50346. E.\,D.\,S. acknowledges 
partial support from NASA LTSARP grant NAG5-9221 and the University of 
Minnesota. This research has made use of the NASA/IPAC Extragalactic Database 
(NED) which is operated by the Jet Propulsion Laboratory, California Institute 
of Technology, under contract with the National Aeronautics and Space 
Administration, and NASA's Astrophysics Data System. 



\clearpage
\end{document}